\documentclass[aps,prl,reprint,superscriptaddress,floatfix,amsmath,amssymb]{revtex4-2}
\usepackage[dvipdfmx]{graphicx}
\usepackage{xcolor}
\graphicspath{{./SU(N)spin/}}
\begin{document}
\newcommand{\ket}[1]
{|#1 \rangle}
\newcommand{\bra}[1]
{\langle #1 |}
\newcommand{\singlet}[0]
{{}^{1}S_{0}}
\newcommand{\triplet}[0]
{{}^{3}P_{2}}
\newcommand{\gup}[0]
{g_{\uparrow}}
\newcommand{\gdown}[0]
{g_{\downarrow}}
\newcommand{\gnd}[0]
{g}
\newcommand{\gupdown}[0]
{g_{\uparrow,\downarrow}}
\newcommand{\exc}[0]
{e}
\newcommand{\kb}[0]
{k_{\rm B}}
\newcommand{\vect}[1]
{\mbox{\boldmath $#1$}}

\title{Observation of antiferromagnetic correlations in an ultracold SU($N$) Hubbard model}
\author{Shintaro Taie}
\altaffiliation{Electronic address: taie@scphys.kyoto-u.ac.jp}
\affiliation{Department of Physics, Graduate School of Science, Kyoto University, Kyoto 606-8502, Japan}
\author{Eduardo Ibarra-Garc\'{i}a-Padilla}
\affiliation{Department of Physics and Astronomy, Rice University, Houston, Texas 77005, USA}
\author{Naoki Nishizawa}
\affiliation{Department of Physics, Graduate School of Science, Kyoto University, Kyoto 606-8502, Japan}
\author{Yosuke Takasu}
\affiliation{Department of Physics, Graduate School of Science, Kyoto University, Kyoto 606-8502, Japan}
\author{Yoshihito Kuno}
\affiliation{Department of Physics, University of Tsukuba, Tsukuba, Ibaraki 305-8571, Japan}
\author{Hao-Tian Wei}
\affiliation{Department of Physics and Astronomy, Rice University, Houston, Texas 77005, USA}
\affiliation{Department of Physics, Fudan University, Shanghai 200433, China}
\author{Richard T. Scalettar}
\affiliation{Department of Physics, University of California, Davis, California 95616, USA}
\author{Kaden R. A. Hazzard}
\affiliation{Department of Physics and Astronomy, Rice University, Houston, Texas 77005, USA}
\author{Yoshiro Takahashi}
\affiliation{Department of Physics, Graduate School of Science, Kyoto University, Kyoto 606-8502, Japan}
\date{\today}

\maketitle
\section*{abstract}
Mott insulators are paradigms of strongly correlated physics, giving rise to phases of matter with novel and hard-to-explain properties. Extending the typical SU(2) symmetry of Mott insulators to SU($N$) is predicted to give exotic quantum magnetism at low temperatures, but understanding the effect of strong quantum fluctuations for large $N$ remains an open challenge. In this work, we experimentally observe nearest-neighbor spin correlations in the SU(6) Hubbard model realized by ytterbium atoms in optical lattices. We study one-dimensional, two-dimensional square, and three-dimensional cubic lattice geometries. The measured SU(6) spin correlations are dramatically enhanced compared to the SU(2) correlations, due to strong Pomeranchuk cooling. We also present numerical calculations based on exact diagonalization and determinantal quantum Monte Carlo. The experimental data for a one-dimensional lattice agree with theory, without any fitting parameters. The detailed comparison between theory and experiment allows us to infer from the measured correlations a lowest temperature of $\left[{0.096 \pm 0.054 \, \rm{(theory)} \pm 0.030 \, \rm{(experiment)}}\right]/\kb$ times the tunneling amplitude. For two- and three-dimensional lattices, experiments reach entropies below where our calculations converge, highlighting the experiments as quantum simulations. These results open the door for the study of long-sought SU($N$) quantum magnetism.
\section*{Introduction}
A recurring question in many-body quantum systems is how the competition of kinetic and interaction energies determines ground state quantum phases. The quantum fluctuations play an essential role in determining the ground state spin structure, which may differ drastically from the mean-field prediction. The SU(2) Hubbard model has long been a prototypical model in which to study these effects, and Hubbard models with an enlarged SU($N$) symmetry have attracted great interest.


The study of SU($N$) quantum magnetism historically originated from the mathematical technique of large-$N$ expansions \cite{Read1983,Affleck1985,Bickers1987,Auerbach2012}. More recently, understanding $N>2$ systems has attracted broader interest, due to the expectation that such systems will display a wide array of exotic physics~\cite{Toth2010,Bauer2012,Nataf2014,Corboz2011,Hermele2011,Romen2020,Yamamoto2020}.
Although $N$ can be large, quantum fluctuations remain important since SU($N$) symmetry prevents spins from becoming classical~\cite{Wu2006,Auerbach2012}.

Although theoretical models with SU($N$) symmetry also have been discussed in connection with real physical systems such as transition metal metal oxides~\cite{Li1998,Tokura2000} and graphene's SU(4) spin-valley symmetry~\cite{Goerbig2011}, the introduction of the symmetry is just a rough approximation.
In contrast, an intrinsic SU($N$) nuclear spin symmetry \cite{Wu2006,Cazalilla2009,Gorshkov2010,Cazalilla2014} is realized in fermionic isotopes of alkaline-earth-metal-like atoms (AEAs), providing unique opportunities for quantum simulation experiments of the SU($N$) Fermi-Hubbard Model (FHM)  \cite{Affleck1988,Honerkamp2004,Assaad2005,Hermele2009}. The SU($N=2I+1$) FHM can be implemented by loading an AEA with nuclear spin $I$ in an optical lattice. This model is given by the Hamiltonian 
\begin{equation}
	H = -t \sum_{\langle i,j \rangle, \sigma} c^{\dagger}_{i\sigma}c^{\phantom{\dagger}}_{j\sigma} + \frac{U}{2} \sum_{i, \sigma \neq \tau} n_\sigma(i) n_\tau(i) - \mu \sum_{i, \sigma} n_\sigma(i), \label{eq_Hamiltonian}
\end{equation}
where $c^{\phantom{\dagger}}_{i\sigma}$ ($c^{\dagger}_{i\sigma}$) denotes the fermionic annihilation (creation) operator for site $i$, $n_\sigma(i) = c^{\dagger}_{i\sigma}c^{\phantom{\dagger}}_{i\sigma}$ is the number operator and $\mu$ is the chemical potential that controls the density. The flavor index $\sigma$ labels the projection quantum number of the nuclear spin $m_I$. Here we employ ${}^{173}$Yb, and $m_I$  is $-5/2,-3/2,\ldots, +5/2$. The tunneling amplitude $t$ and the on-site interaction $U$ do not depend on $\sigma$, giving rise to the SU($N$) symmetry. 

\begin{figure*}[htbp!]	
	\includegraphics[width=\linewidth]{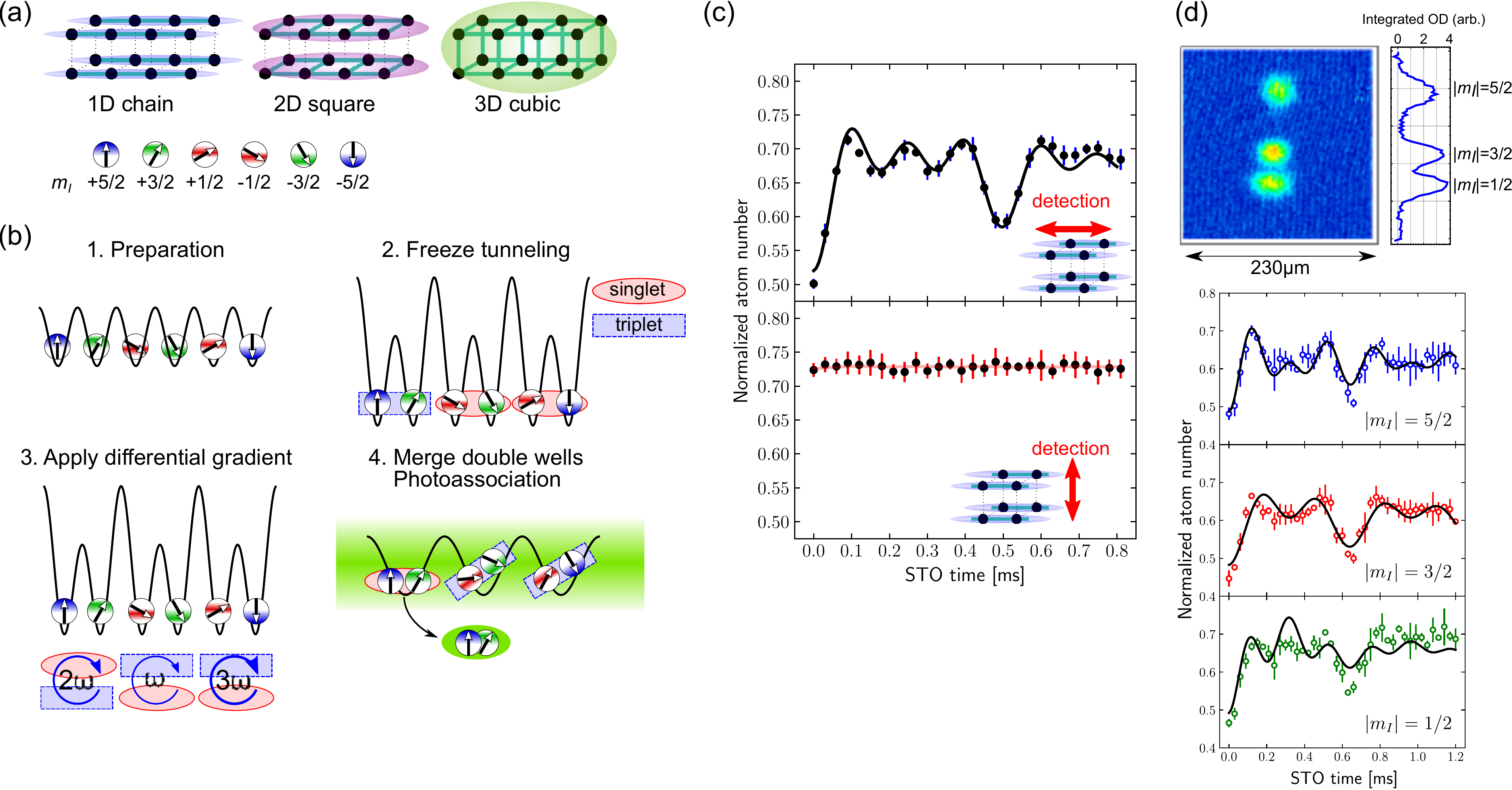}
	\caption{\textbf{Experimental setup.} (a) SU(6) Hubbard systems realized in various configurations of a 3D optical lattice. Spin components are labeled by the nuclear spin projection quantum number $m_I$. (b) Schematic of the experiment. After preparing the equilibrium state and freezing all the tunneling processes, a spin-dependent potential gradient is applied to drive STOs. Subsequently, every two adjacent lattice sites are merged into single sites of the detection lattice, followed by photoassociation which removes atom pairs in antisymmetric spin states. (c) Typical examples of SU(6) STO signal measured for the 1D chain lattice.	The spin-correlation signal for the nearest-neighbors along the chain axis and that along the inter-chain direction are shown in the upper and lower graphs, respectively. Atom numbers are normalized by the total atom number without molecular association processes, and the deviation from unity represents the fraction of singlet states at each time. The initial entropy per particle is $1.45 \kb \pm 0.05 \kb$, and the interaction strength is $U/t = 15.3$. The error bars represent the standard deviation for the 6 independent measurements. (d) STO measurement with optical Stern-Gerlach (OSG) spin separation. Top: Absorption image of the OSG experiment. The image is taken after 5ms time-of-flight. Bottom: Time evolution of the spin population during STO. The solid lines are the fits with the two-frequency model in Eq.~\eqref{eq_OSGfit}. A spin imbalance of 6\% is evaluated from the standard deviation of the atom number in each separated cloud. The error bars represent the standard error of the mean for the 6 independent measurements.} 
	\label{fig::schem}
	\end{figure*}

An important characterization of strongly correlated states is provided by their spin correlation functions. For the SU(2) Hubbard model, antiferromagnetic (AFM) correlations were first observed in dimerized lattices~\cite{Greif2013}, in uniform three-dimensional (3D) lattices using Bragg spectroscopy~\cite{Hart2015}, and in one- and two-dimensional (1D and 2D) lattices using quantum gas microscopy~\cite{Boll2016,Mazurenko2017}. However, correlations in a uniform SU($N$) Hubbard model have not been previously observed. In this work, we observe the nearest-neighbor AFM spin-correlations in an SU(6) ${}^{173}$Yb Fermi gas loaded in 1D, 2D, and 3D optical lattices, and measure them as a function of initial entropy in a harmonic trap. These experimental results are compared with the theoretical calculations with no fitting parameters.

\section*{Experimental setup}
Figure~\ref{fig::schem}(a) depicts our physical system. The SU(6) Fermi gas of ${}^{173}$Yb with atom number $N_{\mathrm{ptcl}} = 2.4(1) \times 10^4$ is adiabatically loaded into 1D chain, 2D square, and 3D cubic lattices that are constructed by a primary optical lattice operating at $532$ nm (see Methods). The 1D chain and 2D square lattices are created by introducing strong tunneling anisotropy into the cubic lattice. The inter-lattice tunneling is less than 5\% of intra-lattice tunneling $t$, and is much smaller than the other energy scales in the system. In our previous work \cite{Ozawa2018} we measured the spin correlation of SU(4) fermions loaded into a double-well system in which the nearest-neighbor correlation is artificially enhanced by strong dimerization. However, in the present work, the SU(6) fermions are loaded into uniform lattices in 1D, 2D, and 3D in which there is no trivial enhancement of spin correlations due to dimerization.

 One can utilize the technique of singlet-triplet oscillations (STO) \cite{Trotzky2010,Greif2013} in an optical superlattice to measure the nearest-neighbor correlations, including for SU($N$) Fermi gases~\cite{Ozawa2018}. The principle of the STO measurement is illustrated in Fig.~\ref{fig::schem}(b). Tunneling is frozen except between pairs of adjacent lattice sites along the measurement axis, which are merged into single sites of a detection lattice which has twice the lattice spacing. Here we utilize the fact that $s$-wave photoassociation (PA) only associates pairs of atoms with a spatially symmetric wavefunction, and thus is only sensitive to spin antisymmetric states in each detection site since the total wavefunction is antisymmetric. Associated molecules quickly escape from the trap, resulting in atom loss. Application of a spin dependent potential gradient before the merging process drives oscillations of spin symmetry, enabling us to also detect spin symmetric states. In this way, we measure the fraction of both ``singlet'' and ``triplet'' states formed within nearest-neighbor lattice sites. The detected SU($N$) counterpart of the SU(2) double-well singlet is a
$\binom{N}{2}$-fold multiplet with the form $(\ket{\sigma, \tau} - \ket{\tau, \sigma})/\sqrt{2}$ ($\sigma \neq \tau$ represents one of the $N$ flavors). Similarly, the double-well triplet is extended to a $\left[ \binom{N}{2}+N \right]$-fold multiplet, among which $\binom{N}{2}$ states with the form $(\ket{\sigma, \tau} + \ket{\tau, \sigma})/\sqrt{2}$ ($\sigma \neq \tau$) are detected by our scheme while $\sigma = \tau$ is not. In the following, we represent the fraction of atoms forming these ``singlet'' and detectable ``triplet'' by $P_s$ and $P_{t0}$, respectively. These are not to be confused with SU($N$) singlets and triplets, which are $N$-body entangled states \cite{Li1998}. The STO measurement is valid only if the contribution from multiple occupancies can be neglected. For that reason, we set the central density to unit filling and the interaction to be sufficiently strong to suppress double occupancies in the primary lattice.

As a measure for the nearest-neighbor spin correlation, we consider a singlet-triplet imbalance defined as
\begin{equation}
I = \frac{P_s - P_{t0}}{P_s + P_{t0}}.
\end{equation}
In addition, we consider a normalized STO amplitude
\begin{equation}
A = P_s - P_{t0}
\end{equation} 
as an alternative measure. As long as the SU($N$) symmetry holds, $A$ is directly related to the SU($N$) spin correlation function $C_{NN}$ \cite{Manmana2011} (see Methods),
\begin{equation}\label{eq::CNN}
C_{NN} = \sum_{\sigma \neq \tau} \bigg[ \langle n_{\sigma}(i) n_{\sigma}(i+1)\rangle - \langle n_{\sigma}(i)n_{\tau}(i+1) \rangle \bigg],
\end{equation}
where $n_\sigma(i+1)$ is a shorthand we use throughout for number operators at a nearest-neighbor of $i$. In the trap, $A$ will be significantly reduced compared to the uniform unit filled case because of the low density at the edge of the sample, and incorporating the effect of the harmonic confinement is important to compare calculations of $A$ with experiments.

Figure~\ref{fig::schem}(c) shows a typical STO signal measured in a 1D chain lattice. To create the spin dependent potential gradient, we utilize linearly polarized laser light close to the ${}^1S_0 \rightarrow {}^3P_1$ resonant frequency (see Methods). As a result, STOs are driven for the spin pairs with different $|m_I|$ ($= 1/2$, $3/2$, and $5/2$), resulting in three different STO frequencies. The ratio of these frequencies $\displaystyle \omega_{\frac{1}{2}-\frac{3}{2}}:\omega_{\frac{3}{2}-\frac{5}{2}}:\omega_{\frac{5}{2}-\frac{1}{2}} = 1:2:3$ is determined by the  Clebsch-Gordan coefficients and does not depend on detuning (see Methods). We analyze the STO signal assuming the SU($N$) symmetry, namely, that all spin combinations equally contribute to correlations. Along the chain axis, we obtain a singlet-triplet imbalance of $I=0.674 \pm 0.052$, indicating the large AFM correlation ($C_{NN}<0$). On the other hand, correlations between chains are zero within the error bar ($I = 0.01 \pm 0.01$) as expected from the negligible inter-chain tunneling.

To verify the expected SU(6) symmetry, we observe the time evolution of each nuclear spin component during STOs. After the standard STO process (driving STO, merging double-wells, and applying PA), lattice potentials are adiabatically ramped down in 6 ms to suppress momentum spread. Then we turn off the optical trap, followed by the application of the optical Stern-Gerlach (OSG) beam for 0.2 ms. The OSG light source is identical with that for the gradient beam for driving STO, with nearly 3 times higher intensity. Therefore the OSG beam is $\pi$-polarized and distinguishes only spin components with different $|m_I|$. Figure~\ref{fig::schem}(d) shows analysis of the spin distribution. The behavior is well reproduced by the two-frequency model (see Methods), indicating that the STO scheme is working as designed. 
\section*{Results}

\begin{figure*}[hbtp!]
\centering
\includegraphics[width=0.8\linewidth]{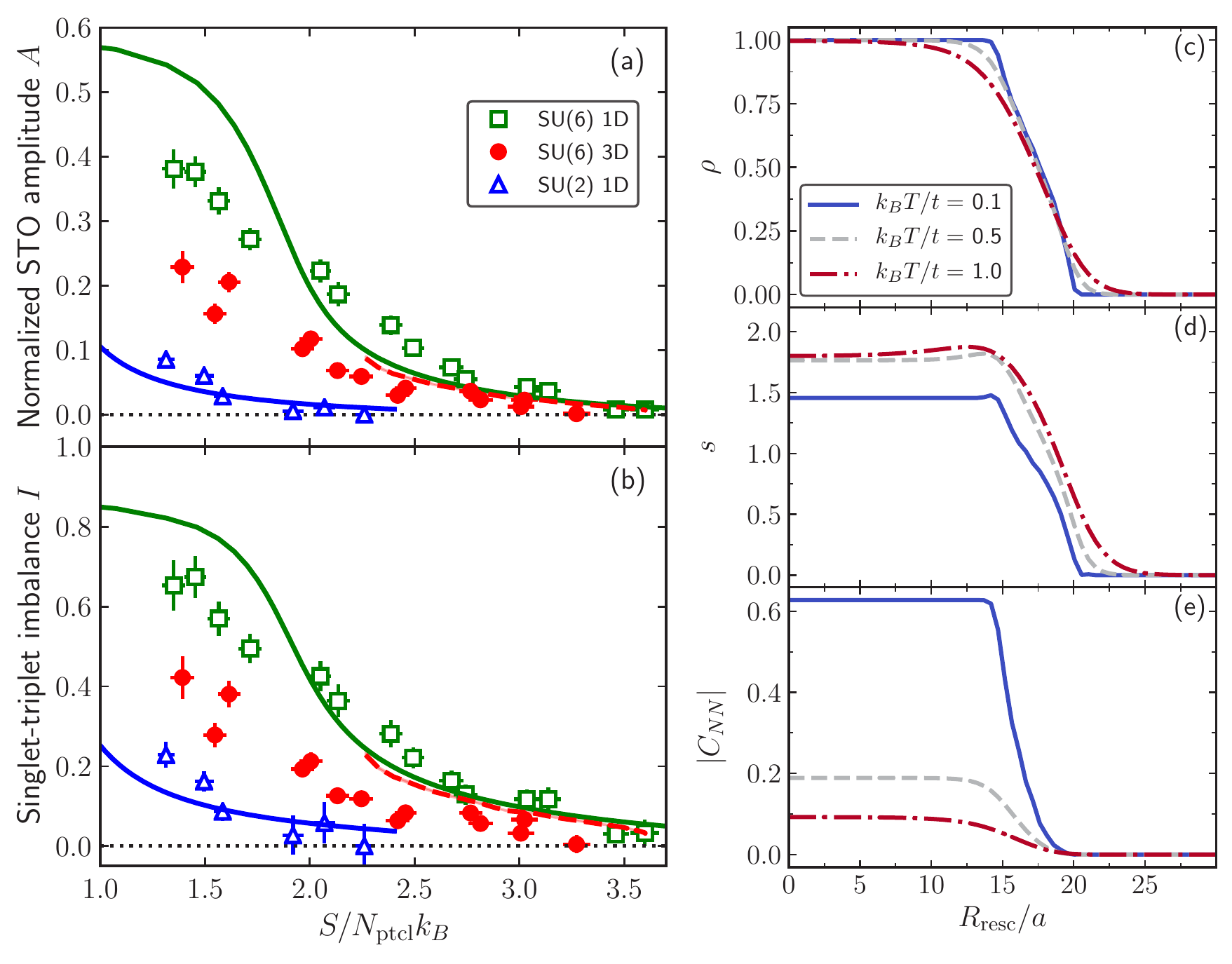}
\caption{\textbf{Entropy dependence of the nearest-neighbor correlations of the SU($N$) FHM at $U/t = 15.3$.} Entropy dependence of (a) the normalized STO amplitude $A$ and (b) the singlet-triplet imbalance $I$ in 1D and 3D lattices are shown. Green squares, red circles, and blue triangles represent the experimental data for SU(6) 1D, SU(6) 3D, and SU(2) 1D systems, respectively. Solid (dashed) lines are the result of ED (DQMC) calculations. The horizontal error bars represent standard error of the mean of the 10 entropy measurements, and the vertical error bars are extracted from the fitting errors in the analysis of the STO signal. (c)-(e) 
Calculated observables as a function of distance to the center of the trap for the 1D system  with trap parameters as in the experiments. (c) Particle number per site, (d) entropy per site, and  (e) nearest-neighbor spin correlation per site as a function of $R_{\mathrm{resc}}/a$, where $R_{\mathrm{resc}} = \sqrt{ \sum_{a=x,y,z} (\omega_a r_a / \overline{\omega})^2}$, $a=266$ nm is the lattice constant, and $\overline{\omega} = (\omega_x \omega_y \omega_z)^{1/3}$ is the geometric mean of the trapping frequencies, for $N=6$ at $U/t=15.3$ in an $L=8$ site chain at $k_BT/t=0.1$, $0.5$, and $1.0$. These temperatures correspond to $S/{N_{\mathrm{ptcl}}k_B = 1.75, \, 2.17,}$ and 2.54, respectively.}\label{fig::IAvsS}
\end{figure*}  

\subsection*{Antiferromagnetic nearest-neighbor spin correlations}

Figures~\ref{fig::IAvsS}(a) and (b) show the nearest-neighbor correlations for 1D and 3D lattices as a function of entropy per particle, with a dramatic enhancement of SU(6) spin correlations compared to the SU(2) correlations in the 1D system. The total entropy $S$ is inferred from a time-of-flight measurement of the weakly interacting gas before lattice loading. The on-site interaction is set to $U/t = 15.3$ for all lattice configurations. In this strongly interacting regime, an important scale is the maximum spin entropy per particle for a singly-occupied site, given by $s_{\rm spin}^{(N)} = \kb \ln N$. Na\"{i}vely, ignoring the spatial inhomogeneity of the trap, a sample with ${S/N_{\mathrm{ptcl}} < s_{\rm spin}^{(N)}}$ is expected to reach the temperature regime where the spin-correlations emerge. For SU(6), $s_{\rm spin}^{(6)} = 1.79 \kb$, while for SU(2), $s_{\rm spin}^{(2)} = 0.69 \kb$, and  $N=6$  systems are therefore expected to show significantly enhanced correlations~\cite{Taie2012,Bonnes2012,Messio2012}.
Our microscopic theory confirms this simple picture, and the observed data show reasonable agreement with theoretical predictions by exact diagonalization (ED) for 1D and determinantal quantum Monte Carlo (DQMC) for 3D, without any fitting parameters.


Figures~\ref{fig::IAvsS}(c)-(e) show theoretically calculated trap profiles of atom number, entropy, and nearest-neighbor spin correlations per site for a 1D system. A rigid Mott plateau is well-developed at ${\kb T/t \sim 0.5}$, and spin correlation rapidly develops for lower temperature. Estimation of the temperature obtained in our experiment is discussed in the next section.

\begin{figure*}[htbp!]
\centering
\includegraphics[width=\linewidth]{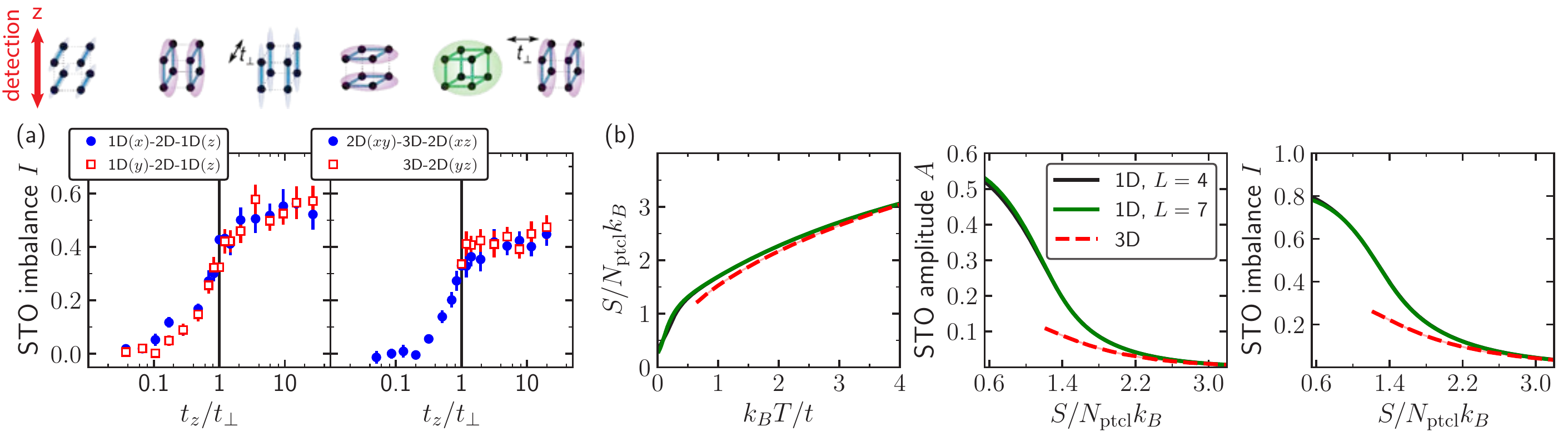}
\caption{\textbf{Dimension dependence of the spin correlations.}
	(a) Left: Spin correlations of an SU(6) Fermi gas as the lattice dimensionality is tuned by lattice anisotropy from 1D to 2D. 
	Correlations along the $z$ axis are measured. Lattices are deformed from $x$-chains (blue circles) and $y$-chains (red squares) to $z$-chains with $U/t = 15.3$, via the 2D square lattice with same $U/t$. Right: Measurement in 2D-3D crossover. Three possible 2D square lattices are connected via the isotropic 3D cubic lattice at $U/t = 15.3$. The initial entropy is $S/{N_{\mathrm{ptcl}}k_B = 1.4 \pm 0.1}$ for both experiments.
	(b) Entropy per particle, normalized STO amplitude, and singlet-triplet imbalance for $N=3$ in $L$-site chains and a $4 \times 4 \times 4$ cubic lattice for $U/t=8$. Note that the the results for 1D $L=4$ and $L=7$ are nearly identical. }\label{fig::dimensionality}
\end{figure*}
\subsection*{Extracting temperature in an optical lattice by theory-experiment comparison}

The present experiments cannot directly measure the temperature at the very low entropies studied here. However, for the 1D systems, the temperature can be inferred by comparing experiment and theory. In 1D, the lowest temperature achieved in the experiments is ${k_BT/t = 0.096 \pm 0.054 \pm 0.030}$, obtained from the experimentally-measured singlet-triplet imbalance $I$ at $S/N_{\mathrm{ptcl}}\kb = 1.45 \pm 0.05$ (see Fig. \ref{fig::1D_thermometry} in Methods section).
The first error bar is an estimate of the finite-size error given by the difference between the finite-size extrapolation to the thermodynamic limit and the 8-site result. The second error bar comes from the experimental uncertainty on the correlations. This is lower than the state-of-the-art temperatures reported in cold atom FHM systems \cite{Hart2015,Parsons2016,Cheuk2016,Hofrichter2016,Mazurenko2017}. Estimates based on $A$ rather than $I$ are similar (see Methods). 
The theory-experiment agreement in 1D suggests the reliability of the experiment in higher dimensions where numerics fail and quantum simulation via experiment is crucial.


For comparison, at the same entropy, the SU(2) system is at ${k_B T/t = 1.008 \pm 0.073 \pm 0.001}$, or to obtain the same singlet-triplet imbalance, the SU(2) system should be at ${S/N_{\mathrm{ptcl}}k_B = 0.499 \pm 0.136 \pm 0.120}$.
Since the state-of-the-art averaged entropy per particle for SU(2) experiments with alkali atoms is around $1\kb$ \cite{Mazurenko2017}, this suggests an experimental advantage for SU($N$) systems in obtaining highly correlated states in optical lattices. 

\subsection{Dependence on lattice dimensionality}
In addition to the dependence on $N$, the correlations significantly depend on dimensionality, with the 1D case exhibiting the largest correlations as shown in Figs.~\ref{fig::dimensionality}(a) and (b). This behavior is similar to previous studies in an SU(2) system \cite{Imriska2014,Greif2015,Ibarra2020} and can be understood by considering lower-dimensional systems as limiting cases of an anisotropic cubic lattice. Removing the tunneling in one direction causes the enhancement of spin correlations in the remaining directions.

In Fig.~\ref{fig::dimensionality}(a), we plot the singlet-triplet imbalance measured through 1D-2D and 2D-3D crossovers with the same initial condition. We measure the correlations along the $z$ axis and change the ratio of $t_z$ to the tunneling $t_\perp$ of the initially weak link. At both the maximum and minimum $t_z/t_\perp$ and at the intermediate point $t_z/t_\perp = 1$, we set $U/t_z = 15.3$. Lattice geometry is smoothly changed between the above three points (see also Methods). We find that spin correlations monotonically decrease as  the lattice is deformed from 1D to 2D. For the 2D-3D crossover, the difference is smaller but the trends of decreasing correlations with increasing dimensionality are still visible. The correlation quickly drops for $t_z/t_\perp < 1$ and becomes undetectable, as expected. 


Numerical calculations show a similar trend. Although DQMC has difficulty in obtaining reliable results for 3D systems at the low temperatures where significant correlations develop for ${U/t=15.3}$ and ${N=6}$, it can calculate the properties of 3D systems for ${U/t=8}$ and ${N=3}$ to low temperature where significant correlations develop.  Because we are considering a smaller $N$ we calculate ED results without using the basis state truncation for 1D $L$-sites chains with $L=4$--$7$. Figure \ref{fig::dimensionality}(b) presents the computed entropy per particle and the spin correlations. Although these are not directly the conditions in the experiments, they do show the same trend of correlations decreasing with increasing dimension.


\section*{Discussion}

We find that the measured nearest-neighbor AFM correlations agree broadly with the theory with no fitting parameters for all temperatures in 1D, and at temperatures where converged theoretical results can be obtained in 3D. In our work for 2D and 3D lattices, we have entered the region where converged theoretical calculations are unavailable and quantum simulation manifests its usefulness.

While we successfully demonstrate the lowest temperature achieved in the FHM in our 1D optical lattice experiment, there is still room for reaching even lower temperatures, for example by engineering spatial redistribution of entropy \cite{Mazurenko2017}.

The spin structures measured in this work are limited to the SU(2)-type nearest-neighbor singlets and triplets. In general SU($N$) systems, more nontrivial spin states arise. For example, the SU($N$) singlet given by the fully antisymmetric combination of $N$ spins plays an essential role in SU($N$) antiferromagnets. Probing such multi-spin entanglement will be an important experimental challenge. Measuring the long range correlations is also of interest. 
One of the most important questions that has not been uncovered yet is whether the long-range ordering persists in the SU($N$) system. Measuring long range correlations will be feasible by using a quantum gas microscope with spin-selective detection technique.  State-of-the-art numerical and analytic calculations, with the use of approximations, have proposed a variety of possible ground states such as flavor-ordered patterns and valence bond solids, among others \cite{Toth2010,Bauer2012,Nataf2014,Corboz2011,Hermele2011,Romen2020,Yamamoto2020}. Experiments are now poised to discriminate finite temperature analogs of such proposed states.

\section*{Methods}\label{sec::method}

\subsection*{Sample preparation}
\begin{figure}[htbp]
	\centering
	\includegraphics[width=0.65\linewidth]{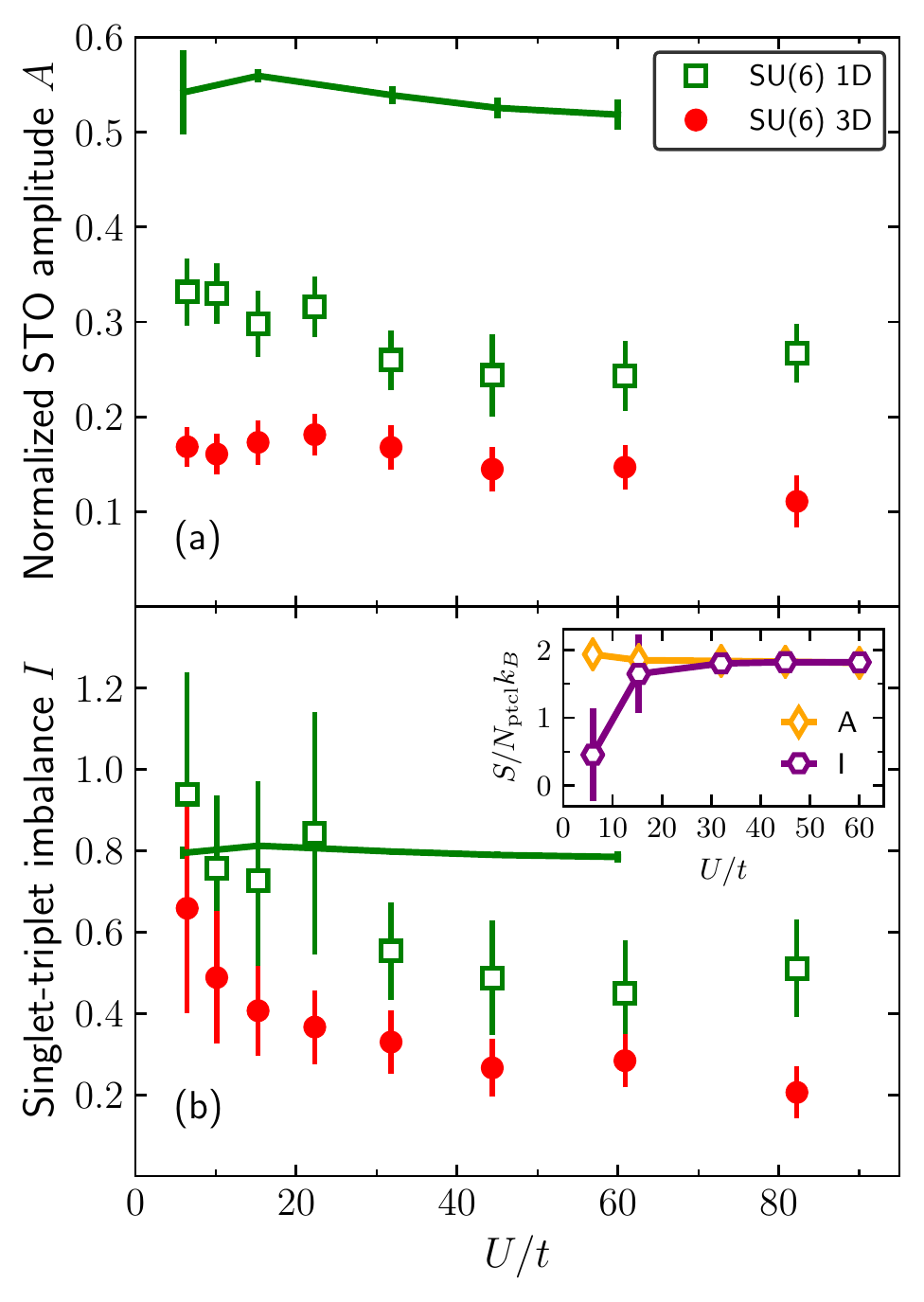}
	\caption{\textbf{Interaction dependence of the nearest neighbor correlations.} Behavior of (a) STO amplitude and (b) singlet-triplet imbalance in 1D and 3D lattices are shown. Experimental data is shown for SU(6) systems with initial entropy $S/N_{\mathrm{ptcl}}\kb = 1.4 \pm 0.1$. The error bars are extracted from the error of fit in the analysis of the STO signal. Solid lines are the result of exact diagonalization calculations for $S/N_{\text{ptcl}}k_B=1.4$, and the error bars correspond to the sum in quadrature of the finite size error and the basis-state truncation error. The inset presents the entropy per particle extracted by fitting it to reproduce the experimentally measured spin correlations. Results saturate at ${S/N_{\mathrm{ptcl}}\kb= 1.818 \pm 0.005}$. Error bars in the inset come from the experimental uncertainty on the correlations.}\label{fig::AIvsU}
\end{figure}

A degenerate Fermi gas of ${}^{173}$Yb is prepared by evaporative cooling in a crossed dipole trap operating at $532$ nm. In the main result obtained in Fig. \ref{fig::IAvsS}, the optical lattices are ramped up to $\vect{s} = (s_x, s_y, s_z) = (7.0, 7.0, 7.0)$ for the 3D cubic lattice, $(6.1, 20.0, 6.1)$ for the 2D ($xz$-) square lattice, and $(20.0, 20.0, 5.0 )$ for the 1D ($z$-)chain lattice. Here, $s_{x,y,z}$ are the lattice depths in units of the recoil energy $E_{\rm R} = \hbar^2(2\pi/\lambda)^2/2m$ with atomic mass $m$ and the wavelength of the lattice laser beams $\lambda = 532$ nm. In the dimensional crossover experiment shown in Fig. \ref{fig::dimensionality}, the lattice geometry is smoothly changed in the form $\vect{s} = (1-p)\vect{s}_1 + p\vect{s}_2\ (0<p<1)$, where $\vect{s}_1$ and $\vect{s}_2$ take the values given above for definite dimensionalities, as well as $(6.1, 6.1, 20)$ for 2D ($xy$-) square and $(5.0, 20.0, 20.0)$ for 1D ($x$-) chain lattices.  The dipole trap together with the optical lattice creates an overall harmonic potential for the sample, whose trap frequencies are $(\omega_{x'}, \omega_{y'},\omega_{z}) = 2\pi \times (102, 44, 155)$ Hz for the 3D lattice, $2\pi \times (105, 49, 158)$ Hz for the 2D lattice, and $2\pi \times (107, 54, 162)$ Hz for the 1D lattice. The principal axes of the trap $x'$ and $y'$ are tilted by $45$ degrees from the lattice axes $x$ and $y$, within the horizontal plane. For the experiment shown in Fig.~\ref{fig::AIvsU}, the trap frequency weakly depends on $U/t$ with the variations within 10\%.

\subsection*{SU(6) singlet-triplet oscillations}
To generate a spin-dependent potential gradient, we apply an optical Stern-Gerlach laser beam close to the ${}^1S_0 \rightarrow {}^3P_1$ resonance. The detuning of $+2.6$ GHz from the $F=5/2 \rightarrow 7/2$ transition is selected to minimize the ratio of the photon scattering rate to the differential light shifts. 
		
The STO signal is analyzed by comparing the total atom number with the number of atoms remaining after removing singlets by photoassociation via the resonance that is located at $-812$~MHz from the ${}^1S_0 \rightarrow {}^3P_1$ ($F=7/2$) transition~\cite{Ozawa2018}.  Assuming that the SU(6) symmetry is not broken, the functional form of the time evolution of remaining atom number is
\begin{equation}
N(t) = -a \exp(-t/\tau) \left[ \cos \omega t + \cos 2\omega t + \cos 3\omega t\right] + b, \label{eq_STOfit}
\end{equation}
with fitting parameters $a$, $b$, $\tau$ and $\omega$. The oscillation frequency $\omega$ is determined from the differential light shift of each spin pair. In general, a differential light shift of a pair $(m_I, m'_I)$ is of the form $\sum_{F'}f(\delta_{F'})[C(F',m_I)-C(F',m'_I)]$, where $f$ is a function of the detuning $\delta_{F'}$ from the excited hyperfine states $F'$ and $C(F',m_I)$ is the transition strength. The constant frequency ratio ($\omega$, $2\omega$, $3\omega$) follows from the fact that, for linear polarization, $[C(F',m_I)-C(F',m'_I)]$ can be reduced to the separated form $C'(F')R(m_I,m'_I)$. Photon scattering and inhomogeneity of the gradient due to the gaussian shape of the OSG beam cause a decay of STO signal, which is described by the exponential decay term in Eq. (\ref{eq_STOfit}). The gradient beam propagates along the $y$ axis and the measurement along the $z$ axis is chosen to suppress the effect of inhomogeneity.

Among the $\binom{6}{2}=15$ spin combinations relevant to STO, linearly polarized light gives rise to the differential light shifts for 12 combinations with different absolute values of $m_I$. The remaining 3 combinations with the same $|m_I|$ do not show STO. Therefore the singlets formed by these pairs are always removed by PA during STOs and the corresponding triplets always remain in the trap. Taking this fact into account, the singlet and triplet fractions in the SU(6) case are expressed as
\begin{eqnarray}
	P_s = \frac{1}{N_{\rm ptcl}}\left[N_{\rm ptcl}- D + 3a - b\right],\\
	P_{t0} = \frac{1}{N_{\rm ptcl}}\left[N_{\rm ptcl}- D - \frac{9a}{2} - b \right].
\end{eqnarray}
where $N_{\rm ptcl}$ is the total atom number without PA and $D$ is the number of atoms on doubly occupied sites (typically less than 3\% of $N_{\rm ptcl}$) which are independently measured without merging and STO processes. Multiple occupancies higher than double are negligibly small.
PA light causes also one-body loss induced by photon scattering, which gives rise to an overestimate of two-body PA loss. In the presence of one-body loss, the substitution $N(t) \rightarrow e^{\gamma \tau_{\rm PA}} N(t)$ is required in analyzing STO, where $\gamma$ is the one-body loss rate and $\tau_{\rm PA}$ is PA pluse duration. In our experiment, $\gamma$ is found to be $0.3$\% of the PA rate and the correction to $N(t)$ is typically 1\%. In the SU(2) case the STO is a simple sinusoid, and the analogous expressions are
\begin{eqnarray}
	P_s = \frac{1}{N_{\rm ptcl}}\left[N_{\rm ptcl}- D + a - b\right],\\
	P_{t0} = \frac{1}{N_{\rm ptcl}}\left[N_{\rm ptcl}- D - a - b \right].
\end{eqnarray}

An atom with specific $|m_I|$ can show STOs with two possible frequencies. With OSG separation, the time evolution of the atom number in each separated cloud $N_{|m_I|}$ is  described by the two-frequency oscillation
\begin{equation}\label{eq_OSGfit}
N_{|m_I|}(t) = -a\exp(-t/\tau) \left[\cos \omega_{|m_I|,1} t + \cos \omega_{|m_I|,2} t\right] + b, 
\end{equation}
with oscillation frequencies
\begin{eqnarray}
(\omega_{|m_I|,1}, \omega_{|m_I|,2}) =
\begin{cases}
(\omega, 3\omega) \ \ \ |m_I|=1/2\\
(\omega, 2\omega) \ \ \ |m_I|=3/2 \\
(2\omega, 3\omega) \ \ \ |m_I|=5/2.
\end{cases}
\end{eqnarray}
Figure \ref{fig::schem}(d) in the main text agrees well with these behaviors of Eq. \eqref{eq_OSGfit}, confirming the validity of the present analysis of STO.  

\subsection*{Numerical calculations for homogeneous systems}
Determinantal Quantum Monte Carlo (DQMC) and exact diagonalization (ED) calculations are used to obtain the values of the thermodynamic quantities, including the density, entropy, and nearest-neighbor spin correlation function for homogeneous systems. These results are used to compute the properties for the trapped system using the local density approximation (LDA), which is described below.

ED results were obtained in $L$-site chains by performing full diagonalization over a reduced Hilbert space (described below) and using finite-size scaling. For computational efficiency, we exploit two aspects of the SU($N$) symmetry. Particle number conservation for each spin flavor
\begin{align}
[N_\sigma,H]=0
\end{align}
with 
 $   N_\sigma = \sum_j  n_{\sigma}(j)$ and the translation symmetries allow us to block-diagonalize the Hamiltonian. Furthermore, we exploit the spin permutation symmetries,
\begin{subequations}\label{eq:spin_perm}
\begin{align}
& \left[S^\sigma_\tau,H\right] = 0 \qquad \forall \sigma,\tau = 1,\dots, N
\end{align}
with
\begin{align}
& S^\sigma_\tau = \sum_i S^\sigma_\tau(i) = \sum_i c_{i \sigma}^{\dagger}c_{i \tau}^{\phantom{\dagger}}
\end{align}
\end{subequations} 
which relate many of the sectors of the Hamiltonian, and therefore one needs to diagonalize only one representative from each sector. 
 
\begin{figure}[htbp!]
	\begin{center}
		\includegraphics[width=0.85\linewidth]{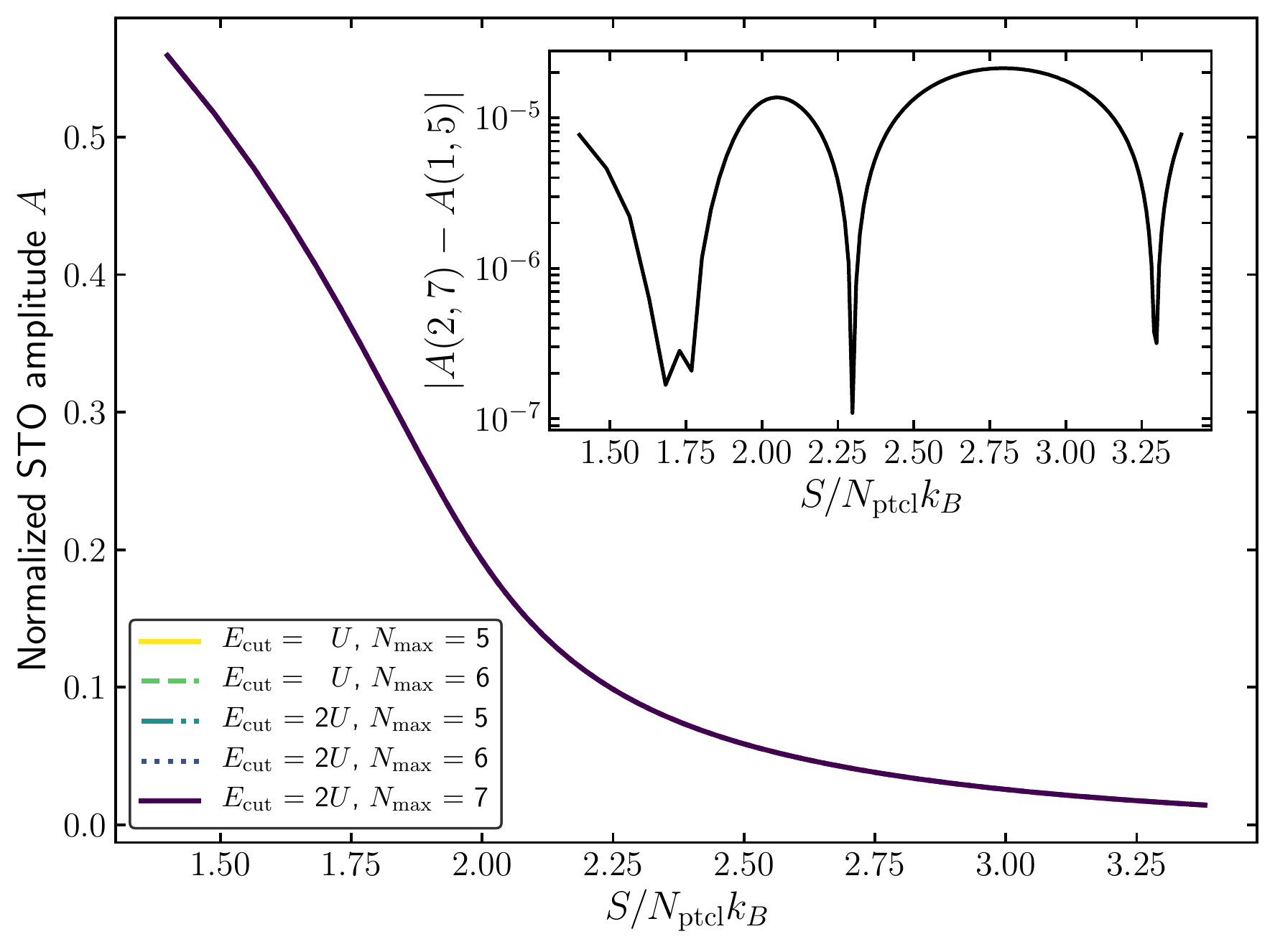}
	\end{center}
	\caption{\textbf{Normalized STO amplitude for an SU(6) Fermi gas in an  $L=5$ site chain with $U/t=15.3$ for different truncations of the Hilbert space.} Basis states with an on-site energy larger than the energy cutoff $E_\mathrm{cut}$, as well as those that exceed the maximum particle number $N_\mathrm{max}$ are disregarded. There is no visible difference between calculations. The inset shows the absolute value of the difference between the $E_\mathrm{cut}=2U,N_\mathrm{max}=7$ and $E_\mathrm{cut}=U,N_{\text{max}}=5$ curves.}\label{fig::1D_truncation_eff}
\end{figure}

In addition to the (exact) symmetries, we employ a basis state truncation, which we systematically converge. First, the Hilbert space only includes states with total particle number lesser or equal to a fixed particle number $N_{\text{max}}$. Second, it omits states if the total on-site energy (the energy associated with the presence of multiple occupancies in the cluster) is larger than $E_{\text{cut}}$. We present results obtained from ${N_{\text{max}}=L+1}$ and ${E_{\text{cut}} =U}$ for $L=5$--$7$ and ${N_{\text{max}}=L}$ for $L=8$.
Figure \ref{fig::1D_truncation_eff} shows that the results for the STO amplitude versus entropy with these truncations are converged to~$\sim 10^{-5}$.

DQMC results for $4\!\times\! 4$ square and $4 \!\times \!4\!\times\! 4$ cubic lattices  were obtained by introducing $N(N-1)/2$ auxiliary Hubbard-Stratonovich fields, one for each interaction term $n_{i \sigma} n_{i \tau}$ \footnote{Previous work applied DQMC to the half-filled SU(2$N$)  FHM using a different, discrete complex Hubbard-Stratonovich decomposition \cite{Wang2014,Zhou2014}}. Following this approach, DQMC calculations for fillings below 1.5 particles per site at ${U/t=15.3}$ can be obtained reliably for temperatures ${k_BT \geq t}$. At lower temperatures, correlation functions become inaccessible to DQMC owing to sign and ergodicity problems. DQMC data were obtained for 5 different random seeds, each with 8000 sweeps through the lattice and the $N(N-1)/2$ auxiliary fields for equilibration and 10000 sweeps for measurements. The inverse temperature was discretized as ${\beta = L \Delta \tau}$ with a Trotter step of ${\Delta \tau = 0.025/t}$. Results are obtained in $\mu-T$ grids with $d\mu = 0.25$ and $dT$ given by the Trotter step for all integers $L \geq 2$. These results are linearly interpolated prior to computing the entropy and using the local density approximation. The entropy per site is computed as the integral of the specific heat, which by thermodynamic relations can be rearranged to
\begin{equation}\label{eq_sup:S_DQMC_1}
s(\mu,T) = N \log(2) + \frac{f(\mu,T)}{T} - \int_T^{\infty} \frac{f(\mu,T')}{{T'}^2} dT',  
\end{equation}
where $f= \epsilon - \mu n$, and $\epsilon$ and $n$ are the energy and particle number per site, respectively. In order to accelerate convergence, we obtain DQMC results up to a temperature cutoff $T_{\mathrm{cut}}$ and use the leading order high temperature series term ($t=0$) in the integral in Eq. (\ref{eq_sup:S_DQMC_1}) for $T>T_{\text{cut}}$. 

\subsection*{Local density approximation} 
Local values of thermodynamic quantities and correlation functions are obtained using the local density approximation (LDA), which replaces intensive observables at a spatial location $\textbf{r}$  with their value in a  homogeneous system with chemical potential ${\mu(\textbf{r}) = \mu_0 -V(\textbf{r})}$, where $\mu_0$ is the global chemical potential  and ${V(\textbf{r})}$ is the external confinement. Applied to the total particle number and to total entropy, this gives 
\begin{align}
{N_{\text{ptcl}}} &= \int\! \frac{d^3r}{a^3} \, n(\mu_0 - V(\textbf{r}),T) \label{eq:LDA}\\
{S} &= \int\! \frac{d^3r}{a^3} \, { s}(\mu_0-V(\textbf{r}),T) \label{eq:LDA-ent}
\end{align} 
where $n/a^3$ and $s/a^3$ are the density and entropy density calculated for the homogeneous system. 
The variables that can be measured experimentally are $N_{\text{ptcl}}$ and $S$ rather than  $\mu_0$ and $T$,  but, given the homogeneous functions $n(\mu,T)$ and $s(\mu,T)$, the $\mu_0$ and $T$  can be obtained from $N_{\text{ptcl}}$ and $S$ by numerically solving Eqs.~\eqref{eq:LDA}-\eqref{eq:LDA-ent}. 

As derived below, the singlet-triplet oscillation (STO) amplitude $A$ and imbalance $I$ are related to the correlation $C_{NN}$, defined in Eq.~(\ref{eq::CNN}), and to the correlation ${\langle n(i)n(i+1) \rangle}$  by 
\begin{eqnarray}
A &= & -\frac{1}{N_{\rm ptcl}} \left[ C_{NN}\right]_{\text{tot}}, \label{eq:A} \\
I &= & \frac{2 A}{\frac{1}{N_{\rm ptcl}} \left[ n(i)n(i+1) \right ]_{\text{tot}} + A }\label{eq:I}
\end{eqnarray}
in the LDA,
where we define
\begin{equation}\label{eq:LDA_obs}
\left[ \mathcal{O}\right]_{\text{tot}}  = \int\! \frac{d^3r}{a^3} \langle \mathcal{O}(\mu(\textbf{r}),T) \rangle.
\end{equation}   

In practice we calculate plots of observable versus $T$ or $S$ as follows. First, we   calculate a list of points $(T,\mu_0)$ where $\mu_0$ is obtained by solving Eq.~(\ref{eq:LDA}) for given $T$ and the   particle number $N_{\text{ptcl}}$ measured in experiment. Then for each such obtained $(T,\mu_0)$, we calculate $S$ and other trap-summed observables of interest. In this way, we plot trap-summed observables as a function of $T$. Details on the grid used and discretization error introduced are given below.

\subsection*{Trap geometry}
\begin{figure}[htbp]
	\centering
	\includegraphics[width=0.85\linewidth]{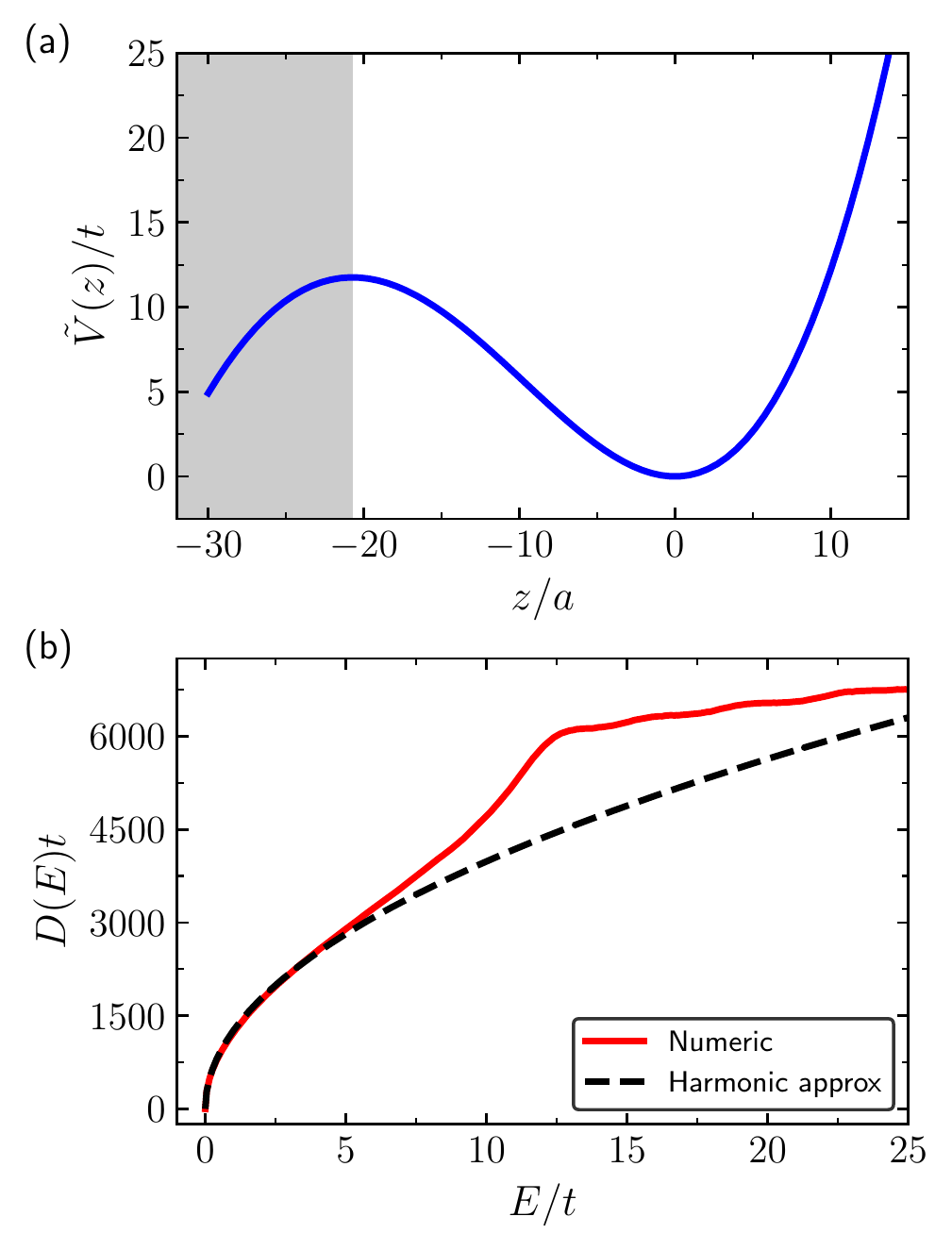}
	\caption{\textbf{Trap anharmonicity.} (a) Full external potential (optical + gravity) profile along the direction of gravity $z$. The shaded region is excluded from the calculation of DOS. (b) Density of states calculated from the external potential for the 3D cubic geometry. The corresponding harmonic approximation is also shown. The atoms are sensitive only to the density of states for $E/t\lesssim 10$.}\label{fig_DOS}
\end{figure}
Due to the large atomic mass of Yb, the effect of gravity is severe for our optical trap, especially in the final stage of evaporative cooling [see Fig.~\ref{fig_DOS} (a)]. To include the anharmonic effect in our LDA calculation, we evaluate the density of state (DOS) defined as
\begin{equation}\label{eq_DOS}
D(E) = \frac{\partial \Sigma}{\partial E}, \quad \Sigma(E) = \sum_{\mathrm{lattice \, sites} \, i} \Theta \left(E - \tilde{V}(\textbf{r}_i)\right),
\end{equation}
where $\tilde{V}(\textbf{r}_i)$ is the full external potential at the site $i$ except the periodic part forming optical lattices.
Note that this DOS function becomes exact only in the atomic limit $t\rightarrow0$, but is always valid for the use in the LDA integral described below. Trap-summed observables are calculated using Eqs. (\ref{eq:LDA}-\ref{eq:LDA_obs}). These integrals over space are then rewritten as integrals over energy with the DOS,
\begin{equation}
\left[ \mathcal{O}\right]_{\text{tot}}  = \int \! \mathrm{dE} \, D(E) \langle \mathcal{O}(\mu_0 - E,T) \rangle.
\end{equation}
In the harmonic approximation, the DOS is given by
\begin{equation}\label{eq_DOSharmonic}
D(E) =  \frac{2\pi}{a^3}\left( \frac{2}{m \bar{\omega}^2} \right)^{3/2} E^{1/2}
\end{equation}
where $\bar{\omega}$ is the geometric mean of the trap frequencies.

In Fig.~\ref{fig_DOS}(b), we plot the evaluated DOS for the 3D cubic lattice. In calculating Eq.~\eqref{eq_DOS}, we exclude the spatial region outside the potential barrier, where $\tilde{V}$ becomes a uniformly decreasing function along the direction of gravity [see Fig. \ref{fig_DOS}(a)]. In the low energy region the DOS is well reproduced by the harmonic approximation. As energy increases, the DOS starts to exceed the harmonic prediction due to the nearly flat potential where the optical potential gradient is competing with the gravitational one. For even higher energies, the DOS falls below the harmonic approximation because the contribution is limited only from the upper half of the trap. The difference between the results calculated in the harmonic approximation and using the full potential is small, never larger than $2.4\times10^{-2}$ for the normalized STO amplitude and imbalance in the range of entropies presented in the main text.

\subsection*{Singlet-triplet oscillation amplitude and imbalance}  
In the limit where there are no multiple occupancies, the populations in the singlet and triplet states $p_{\sigma \tau}^s(i)$ and $p_{\sigma \tau}^{t0}(i)$ for an STO with spin components $\sigma$ and $\tau$ in the dimer located on sites $i$ and $i+1$ are given by the expectation value of the projection operators,
\begin{align}
\widehat{P}_{\sigma \tau}^s(i) = \frac{1}{2} &\bigg( c_{i,\sigma}^\dagger c_{i+1,\tau}^\dagger - c_{i,\tau}^\dagger c_{i+1,\sigma}^\dagger \bigg) \nonumber \\ 
 & \times \bigg( c_{i+1,\tau}^{\phantom{\dagger}}c_{i,\sigma}^{\phantom{\dagger}}  -  c_{i+1,\sigma}^{\phantom{\dagger}} c_{i,\tau}^{\phantom{\dagger}} \bigg), \\
\widehat{P}_{\sigma \tau}^{t0}(i) = \frac{1}{2} &\bigg( c_{i,\sigma}^\dagger c_{i+1,\tau}^\dagger + c_{i,\tau}^\dagger c_{i+1,\sigma}^\dagger \bigg) \nonumber \\
 & \times \bigg( c_{i+1,\tau}^{\phantom{\dagger}}c_{i,\sigma}^{\phantom{\dagger}}  +  c_{i+1,\sigma}^{\phantom{\dagger}} c_{i,\tau}^{\phantom{\dagger}} \bigg).
\end{align}
Note that these refer to SU($2$) singlets involving components $\sigma$ and $\tau$ rather than SU($N$) singlets. It is useful to introduce  spin-1/2 operators $S^z_{\sigma \tau}(i)=[n_\sigma(i) - n_\tau(i)]/2 $ for the pair of states $\sigma$ and $\tau$.
By the SU($N$) symmetry the population difference and sum are equal to \cite{Greif2013}
\begin{align}
p_{\sigma \tau}^s(i) - p_{\sigma \tau}^{t0}(i) =& -4 \langle S^z_{\sigma \tau}(i)S^z_{\sigma \tau}(i+1) \rangle, \\
p_{\sigma \tau}^s(i) + p_{\sigma \tau}^{t0}(i) =& + \frac{1}{2}\langle n(i)n(i+1) \rangle \nonumber \\
 &-2 \langle S^z_{\sigma \tau}(i)S^z_{\sigma \tau}(i+1) \rangle.
\end{align}
The fractions of atoms forming singlets ${P_s= N_s/N_{\rm ptcl}}$ and triplets  ${P_{t0}= N_{t0}/N_{\rm ptcl}}$ are obtained from a sum over each dimer in the lattice, or equivalently $(1/2)\sum_i \cdots$, and all the possible $\sigma$-$\tau$ spin pairs,
\begin{subequations}\label{eq:ps_pt0}
	\begin{align}
		P_s &= \frac{2}{N_{\rm ptcl}} \left[ \frac{1}{2}\sum_i \left(\frac{1}{2}\sum_{\sigma \neq \tau} p_{\sigma \tau}^s(i) \right) \right], \\
		P_{t0} &= \frac{2}{N_{\rm ptcl}} \left[ \frac{1}{2} \sum_i\left( \frac{1}{2}\sum_{\sigma \neq \tau} p_{\sigma \tau}^{t0}(i) \right)\right].
	\end{align}
\end{subequations} 
Therefore the global STO amplitude is
\begin{equation}
A = P_s -P_{t0} = \sum_i \sum_{\sigma \neq \tau} \left[ \frac{-4 \langle S^z_{\sigma \tau}(i)S^z_{\sigma \tau}(i+1) \rangle}{2 N_{\rm ptcl}} \right],
\end{equation} 
which in terms of the $n_\sigma(i)$ is
\begin{align}\label{eq_sup:A}
A &= - \sum_i \sum_{\sigma \neq \tau} \left[ \frac{\langle n_\sigma(i) n_\sigma(i+1) \rangle - \langle n_\sigma(i) n_\tau(i+1) \rangle}{N_{\rm ptcl}} \right] \nonumber \\
& = -\frac{1}{N_{\rm ptcl}} \sum_i C_{NN}(i).
\end{align}
The STO imbalance $I$ is defined as
\begin{equation}
I=\frac{P_s -P_{t0}}{P_s + P_{t0}}
\end{equation}
so
\begin{equation}\label{eq_sup:I}
I = \frac{2 A}{ \frac{1}{N_{\rm ptcl}}  \sum_i \langle n(i)n(i+1) \rangle + A},
\end{equation}
where $\langle n(i)n(i+1) \rangle$ is the density-density correlation function.  Eqs. (\ref{eq_sup:A}-\ref{eq_sup:I}) directly yield Eqs. (\ref{eq:A}) and (\ref{eq:I}).

%
%
\subsection*{Thermometry in 1D} 
\begin{figure*}[htb!]
	\begin{center} \includegraphics[width=\linewidth]{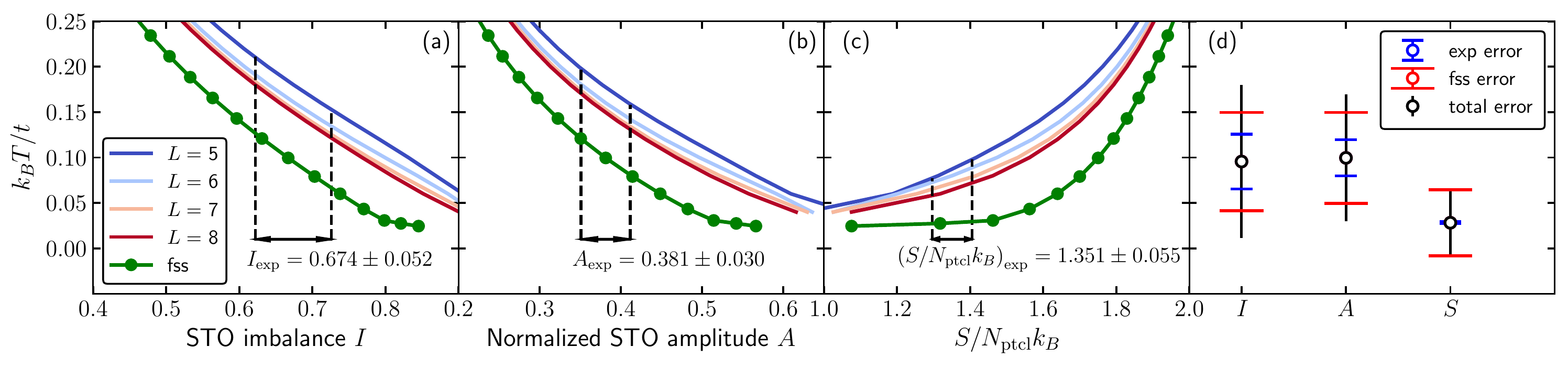}
	\end{center}
	\caption{\textbf{Temperature of a 1D SU(6) Fermi gas at $U/t=15.3$.} The vertical dashed lines indicate the range of the largest experimentally measured STO imbalance in 1D that is consistent with error bars. The temperature of this datapoint is inferred from the finite-size scaling curves and the results are summarized in panel (d).  The fss error is a conservative estimate of the finite-size error, the  difference between the finite-size scaled results and the $L=8$ site chain. The exp error comes from the experimental uncertainty on the correlations.}\label{fig::1D_thermometry}
\end{figure*}

In Fig.~\ref{fig::1D_thermometry} we present how we determine the lowest temperature achieved in the 1D experiments ($k_BT/t = 0.096 \pm 0.054 \pm 0.030$). Estimates based on $A$ rather than $I$ give similar result. The estimate based on the lowest entropy prior to lattice loading predicts somewhat lower temperature, although still consistent within error bars. A small increase in temperature could result from non-adiabatic effects during the lattice loading. In Fig.~\ref{fig::AIvsU}, we show the interaction dependence of the spin correlations. The tendency toward larger discrepancy between theory and measurement with larger interactions (equivalent to deeper lattice depths) suggests that heating is important for deeper lattices. 

\subsection*{Exact diagonalization error estimates}
\begin{figure}[htb!]
	\begin{center}
		\includegraphics[width=\linewidth]{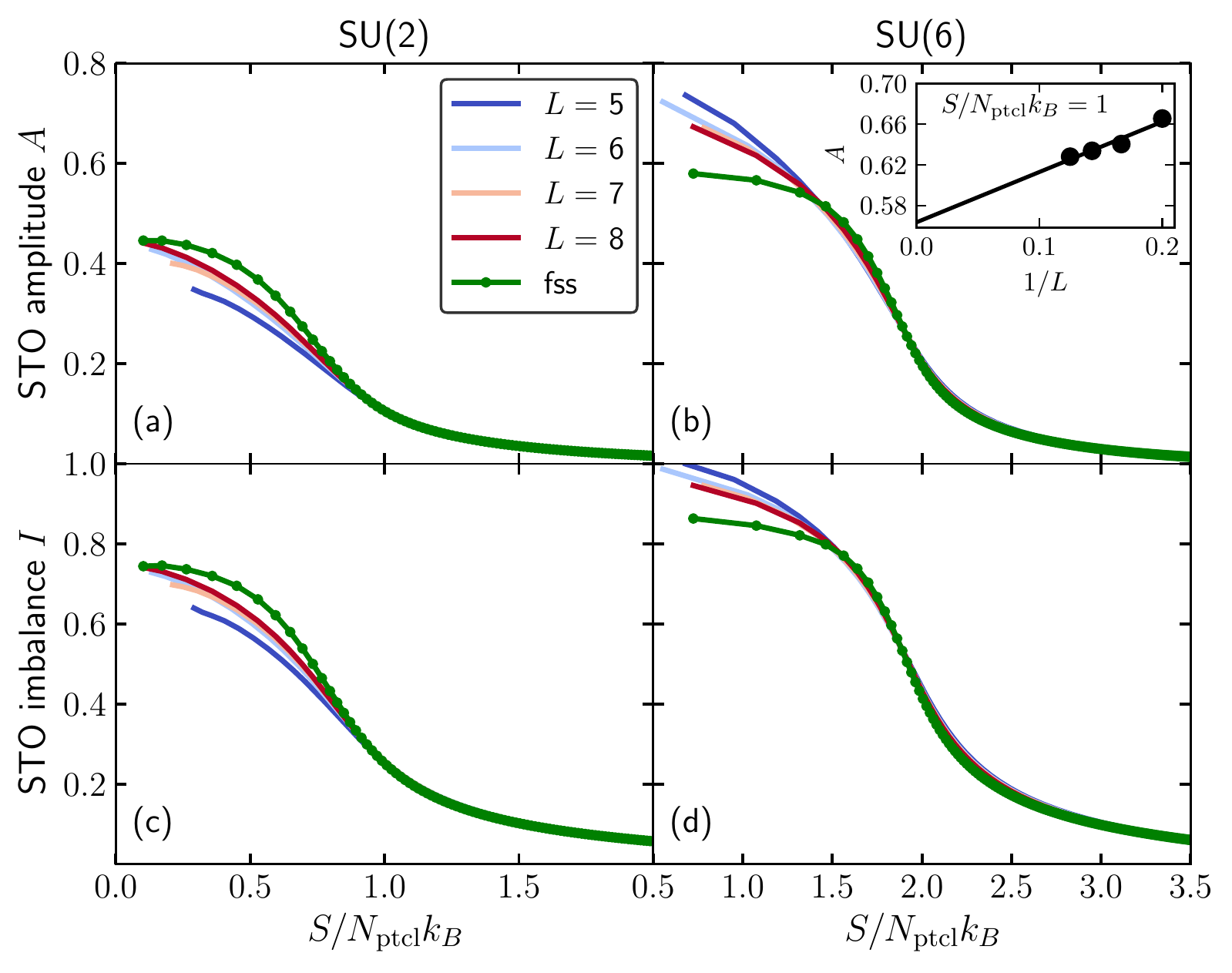}
	\end{center}
	\caption{\textbf{Finite-size scaling in 1D.} Normalized STO amplitude and imbalance for SU(2) and SU(6) Fermi gases in $L$-site chains and the results after finite-size scaling (fss) for $U/t=15.3$. The inset in panel (b) illustrates the finite-size scaling procedure. }\label{fig::1D_finite_size_eff}
\end{figure}

Errors for the exact diagonalization results arise from two sources: finite-size error and truncation of the Hilbert space using the on-site energy and maximum particle number  criteria. Figures~\ref{fig::1D_finite_size_eff} present the normalized STO amplitude and imbalance for SU(2) and SU(6) in 1D for different system sizes $L=5,\ldots, 8$, as well as the finite-size extrapolation. Results in the main text are presented after finite-size scaling at fixed entropy per particle, which is performed by fitting the results for $L=5,\ldots, 8$ to  $\mathcal{O}_L =~ \mathcal{O}_\infty + m/L$ with $\mathcal{O}_\infty$ and $m$ as fitting parameters.  Validity of the Hilbert space truncation is tested by varying the energy cutoff $E_\mathrm{cut}$ as well as the maximum particle number $N_\mathrm{max}$. Figure \ref{fig::1D_truncation_eff} demonstrates that for the interaction strengths considered herein, the truncation is extremely accurate, with no visible differences for any parameters. 

\subsection*{Determinantal Quantum Monte Carlo error estimates}
\begin{table*}[h!]
	\centering
	\begin{tabular}{l | p{2cm} p{2cm} p{2cm} } 
		\hline
		Error source (homogeneous, worst case over all $\mu$ and $T$) & $S$ vs $T$ & $A$ vs $T$ & $I$ vs $T$  \\
		\hline\hline
		Finite $T_{cut}$ & $1.3\times10^{-4}$  & \multicolumn{1}{c}{--} &  \multicolumn{1}{c}{--}  \\
		Statistical  & $1.2\times10^{0}$ & $1.8\times10^{-1}$ & $8.6\times10^{-2}$ \\
		\hline
		Error source (in the trap) & $S$ vs $T$ & $A$ vs $S$ & $I$ vs $S$  \\
		\hline \hline
		Finite size (2D) & $1.8\times10^{-2}$ & $1.1\times10^{-2}$ & $2.4\times10^{-2}$  \\
		Trotter-step & $1.8\times10^{-2}$ & $4.6\times10^{-3}$ & $9.5\times10^{-3}$ \\ 
		$\mu$ grid coarseness & $6.9\times10^{-3}$ & $7.2\times10^{-3}$ & $1.5\times10^{-2}$ \\
		$T$ grid coarseness & $1.4\times10^{-2}$ & $2.4\times10^{-3}$ & $5.3\times10^{-3}$ \\
		Statistical (adiabatic loading) & $1.0\times10^{-2}$ & $1.6\times10^{-3}$ & $3.2\times10^{-3}$ \\
		\hline
	\end{tabular}
	\caption{\textbf{Error estimates for the DQMC calculation at $\vect{U/t = 15.3}$.}
		Errors are presented at $\kb T/t=1$.  
		Most errors decrease  with increasing temperature.}
	\label{table:SU(N) FHM DQMC error bounds U15p3}
\end{table*}

\begin{table*}[h!]
	\centering
	\begin{tabular}{l | p{2cm} p{2cm} p{2cm} } 
		\hline
		Error source & $S$ vs $T$ & $A$ vs $S$ & $I$ vs $S$  \\
		\hline\hline
		Trotter-step & $3.2\times10^{-2}$ & $7.8\times10^{-4}$ & $2.1\times10^{-3}$ \\ 
		$\mu$ grid coarseness & $7.8\times10^{-4}$ & $1.6\times10^{-4}$ & $5.2\times10^{-4}$ \\
		$T$ grid coarseness & $1.3\times10^{-1}$ & $1.8\times10^{-3}$ & $4.9\times10^{-3}$ \\
		Statistical (adiabatic loading) & $2.0\times10^{-4}$ & $5.0\times10^{-5}$ & $1.2\times10^{-4}$ \\
		\hline
	\end{tabular}
	\caption{\textbf{Error estimates for the different error sources involved in the DQMC calculation for $\vect{U/t=8}$}. We report the largest error in the whole range of temperatures/entropies considered. The $T$ grid coarseness error for $S$ vs $T$ monotonically increases from ${2.0 \times 10^{-2}}$ at $\kb T/t = 0.64$ to the value reported in the table, $1.3 \times 10^{-1}$, at $\kb T/t = 4$.}
	\label{table:SU(N) FHM DQMC error bounds U8}
\end{table*}

DQMC results have several sources of error, which are estimated and presented in Table \ref{table:SU(N) FHM DQMC error bounds U15p3} for SU(6) and ${U/t=15.3}$, and for SU(3) at $U/t=8$ in Table \ref{table:SU(N) FHM DQMC error bounds U8}. All error estimates are presented after the adiabatic loading calculation obtained by calculations for a ${4 \times 4 \times 4}$ lattice at ${k_BT/t=1}$ unless explicitly stated otherwise. In this section we will briefly discuss how each error source was estimated.

The largest and most-difficult to quanify error comes from  finite-size effects. As a proxy, we estimate finite-size effects by studying 2D, and taking the difference between  the $4 \times 4$ results and the  $6 \times 6$ results. Larger system sizes, in particular in 3D, remain  inaccessible at present. 

The inverse temperature discretization error is estimated as the difference of the results obtained with Trotter steps $\Delta \tau = 0.025$ and $\Delta \tau = 0.05$. 

The entropy per site $s$ at temperature $T$ is given by Eq. (\ref{eq_sup:S_DQMC_1}). Errors in the calculation of $s$ arise from by the finite value of the temperature cutoff $T_{cut}$. This error was estimated in the homogeneous case by comparing the results obtained with $k_BT_{\text{cut}}/t =500, \,800, \, 1000$. 

Errors in numerical integration procedures such as the local density approximation summing of observables in the trap  and the computation of the entropy are estimated by varying the coarseness of the $\mu$-$T$ integration grids. Such estimations were obtained by coarsening them by a factor of two and comparing the results.

Statistical errors are presented for both the homogeneous case and after adiabatic loading for 5 different random seeds. These errors are presented as the standard error of the mean.

\section{Acknowledgements}
We thank H. Ozawa for contributions in the early stage of the experiment. The experimental work was supported by the Grant-in-Aid for Scientiﬁc Research of JSPS (Nos. JP17H06138, JP18H05405, and JP18H05228), the Impulsing Paradigm Change through Disruptive Technologies (ImPACT) program, JST CREST (No. JPMJCR1673), and MEXT Quantum Leap Flagship Program (MEXT Q-LEAP) Grant No. JPMXS0118069021. Y. K. acknowledges the support of the Grant-in-Aid for JSPS Fellows (No.17J00486).
The work of K.R.A.H., E.I.G.P., and H.W. was supported in part by the Welch Foundation through Grant No. C1872, the Office of Naval Research Grant No. N00014-20-1-2695, and the National Science Foundation through Grant No. PHY1848304. The work of R.T.S. was supported by the grant DE-SC0014671 funded by the U.S. Department of Energy, Office of Science.

\bibliography{SU6_Magnetism}
\end{document}